# A MODIFIED CELLULAR AUTOMATON USING ACTIVATION AND INHIBITION REGIONS GEOMETRICALLY COMPATIBLE WITH BIAXIAL ANISOTROPY

R. PIASECKI[1], W. OLCHAWA[1], D. FRĄCZEK[2]

1  Institute of Physics, University of Opole,

Oleska 48, 45-052 Opole, Poland

2  Department of Materials Physics, Opole University of Technology,

Katowicka 48, 45-061 Opole, Poland

Young's cellular automaton, recently applied to study the spatiotemporal evolution of binary patterns for favorable/hostile environments, has now been modified from a different point of view. In this model, each differentiated cell (DC) produces two diffusing morphogens: a short-range activator and a long-range inhibitor. Their combination creates the so-called local '$w$' field. Undifferentiated cells (UCs) are passive. The question arises how to adapt it to modelling patterning processes in anisotropic substrates with a biaxial dependence of the morphogen diffusion rate. We use activation/inhibition regions with appropriate shape geometry defined by the so-called deformation parameter $p$. We complement this model by adding a physically reasonable transition zone with controlled local field slope. The patterning process uses the morphogenetic field $W$ calculated separately for each cell, which is the sum of the '$w$' values generated by all regional DCs surrounding the cell. It acts as a chemical signal determining the next state of the cell. We also introduce a threshold $W^*$ defining the required absolute chemical signal strength. The state of each cell can change depending on the rules for $W$ and $W^*$. This improves the stability of the model evolution and extends its applications. Using two pairs of two-point orthogonal correlation functions, we reveal their directional sensitivity to changes in biaxial anisotropy. Finally, the general two-parameter dependence of the average final DC concentration on the geometry of the activation/inhibition regions and on the long-range inhibitor value was illustrated. This facilitates the recognition of characteristic features of the evolution of DC concentration in our model.

## 1. Introduction

For reasons of evolution, many different biological patterning processes can occur in nature. A description of various models and mechanisms involved in such processes can be found in the comprehensive book [1]. Since the publication of Turing's work [2], diffusion-based patterning models have been particularly popular; see among others [3-6]. However, the approach using a spatially discrete cellular automaton (CA) on a square grid is particularly convenient in computer simulations [7]. Adapting the CA-type model proposed by Young [8] for vertebrate skin patterning, we analyzed the influence of a favorable, neutral, or hostile but isotropic environment on spatial patterning [9]. A combination of two morphogens, activator and inhibitor, produced exclusively by differentiated cells and diffusing with different speeds creates a so-called local field. However, the fate of a given cell depends on the associated regional morphogenetic field. This field is the sum of the corresponding local field values from the DCs surrounding the cell. Of course, these values depend on the type of region occupying by DCs. Despite their *simple* structure, models of this type reproduce, most often on a square grid, the basic features of vertebrate skin: spots, stripes or mixed forms.

It is worth mentioning here structurally more complex versions of the Turing model. For example, the probabilistic approach to multicolor activation and lateral inhibition [10], the mathematically advanced model taking into account the shape of the activation-inhibition kernel [11], or the totalistic CA model on a hexagonal grid additionally using the rules of the game of life [12], to name just a few.

In this paper, we focus on the specific biaxial diffusivity of chemical signals. Such behavior may be naturally related to the *anisotropic* nature of the substrates. Therefore, we generally use non-circular activation/inhibition regions. In other words, in the modified Young model, we include a local morphological field with a shape geometry consistent with the biaxial anisotropy of the environment. We suggest that this approach may also be useful when the initial configuration contains clusters having biaxial symmetry. On the other hand, adding a transition zone with a controlled slope of the local field and taking into account a certain threshold of the required absolute strength of the cumulative chemical signal makes the model more physical and shows a higher stability, respectively.

The main result of our work reveals a general graphical relationship between the mean size of the final population of DCs and any combination of the two leading parameters. The first describes long-range inhibition, while the second refers to the shape of the activation/inhibition regions associated with biaxial anisotropy. In this way, the characteristic features of the evolution of final concentration can easily be visualized.

The article has the following structure. First, we will present details of the modified model. We then illustrate our approach, using morphologically different examples, as well as an identical initial random population, but with different biaxial substrate anisotropy. Finally, we will make some concluding remarks in the general context.

## 2. Modifications of Young's model

The effect of a favorable, neutral or hostile but *isotropic* environment on the formation of spatial binary patterns was analyzed for the extended Young's cellular automaton in [9]. This model uses diffusible morphogens, a short-range activator $w_1 > 0$ and a long-range inhibitor $w_2 < 0$. Recall that these are produced exclusively by differentiated cells (DCs), shown here as black pixels.

The remaining cells, called undifferentiated cells (UC), are passive in this context and are represented by white pixels. The combination of these two morphogens creates a stepped local activation/inhibition field, with two fixed values: positive in the inner region and negative in the outer one. Despite its approximation, Young's model provides an acceptable description of the process of formation of complex binary forms [8, 9].

Still, useful modifications of the model can be considered. Note that the controlled slope of the local morphological field in the additional transition zone proposed in this paper improves Young's CA. This approach is physically more acceptable, although still simplified. Fig. 1 shows a partial section of the local field along the x-semiaxis. The meaning of other symbols is given below in the text. On the other hand, possible modifications of the shape of activation/inhibition areas would allow extending the scope of modelling isotropic environments to some kind of anisotropic ones with the specific direction-dependent diffusion of morphogens. In Ref. [8], a related example can be found, where replacement of the circular activation/inhibition area with elliptical one enabled the generation of strip structures imitating uniaxial anisotropy. The cellular automaton with activation/inhibition regions of appropriate geometry can also be used when the initial configuration includes compact clusters with shapes defined, for example, by the deformation parameter $p$ [13]. In such cases, our approach seems to be the natural choice.

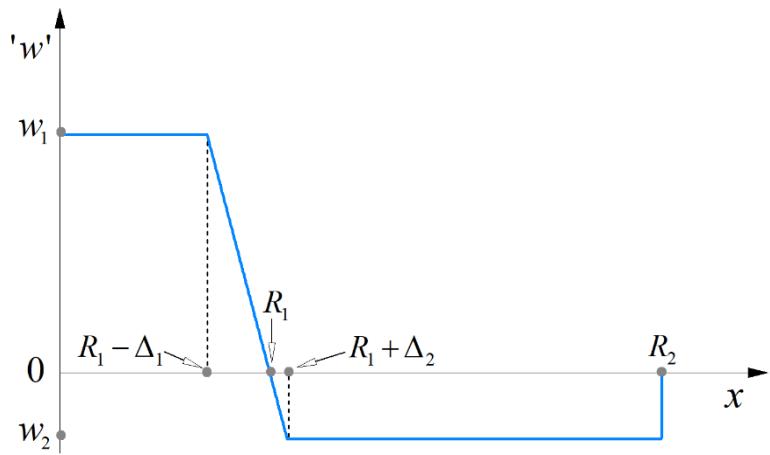

Fig. 1. Illustration of the modification of the activation-inhibition field for Young's model [8]. Only a partial cross-section of the local field along the x-axis is shown (a similar diagram corresponds to the y-axis). Note that the proportion $w_1/\Delta_1 \equiv |w_2|/\Delta_2$ relates to the slope of the local field in the transition zone.

In the modified model, for simulation purposes, we first use a smaller square grid of linear size L = 81 pixels with periodic boundary conditions (PBC) in both directions. According to the rules given below, any initial random distribution of a given percent of differentiated (black) cells in an array of undifferentiated (white) cells can evolve into a binary pattern. For illustration and testing purposes, we will also use synthetic configurations for DCs. Since the emphasis is on biaxial anisotropy, a single deformation parameter $p$ [13] is preferred to define the shape of the areas, as it is particularly easy to use. Each cell at a discrete position $(x_0, y_0)$ is influenced by the sum of morphogens from all $i$-th DC$(x_i, y_i)$ occupying activation/inhibition regions bounded by a simplified super-ellipse with semi-axis $r$ = semi-major axis = semi-minor axis, which is given by the formula

$$\left| x_i - x_0 \right|^{2p} + \left| y_i - y_0 \right|^{2p} = r^{2p} \tag{1}$$

Within this approach, the deformation parameter $p \geq 0$ indicates to what extent the shape of the region has deformed from that of a circle ($p = 1$). For simplicity, further in the text we omit the sub-index '$i$' except where necessary. Some examples of specifically shaped regions placed around point $(0, 0)$ are given in Fig. 2.

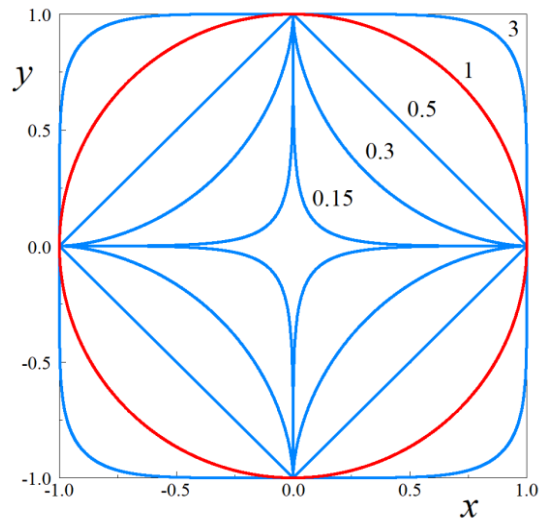

Fig. 2. Examples of the shapes of the region for semi-axis $r = 1$ using different values of the deformation parameter $p$. The circular case with $p = 1$ is marked in red. Analytically, at the limit $p \rightarrow \infty$ the non-circular area is a perfect square.

Using a modified local activation/inhibitory field, instead of the two regions, that is an internal activating type (I) and an external inhibitory type (III), an additional intermediate activation/inhibitory type (II) region is considered. The basic parameters of the model are $w_1$, $w_2$, $R_1$, $R_2$, $\Delta_1$ and $p$. Note that we treat parameter $\Delta_2$ as auxiliary because it is defined by the relation $\Delta_2 \equiv (|w_2|/w_1)\Delta_1$. To perform simulations for the selected values of parameter $p$, the exemplary model parameters were set: $R_1 = 4.8$, $R_2 = 10$ and $\Delta_1 = 1.8$.

The modified patterns formation mechanism using the local morphogenetic field '$w$' includes:

a) constant short-range activation $w_1 > 0$ in the inner region (I) for $1 \leq r \leq R_1 - \Delta_1$,

b) linearly decreasing activation/inhibition according to the formula $w_{12}(r) = (-w_1/\Delta_1)\, r + w_1 R_1/\Delta_1$ with the simple identity $w_1/\Delta_1 \equiv |w_2|/\Delta_2$ in the transitional region (II) for $R_1 - \Delta_1 < r \leq R_1 + \Delta_2$,

c) constant long-range inhibition $w_2 < 0$ in the outer region (III) for $R_1 + \Delta_2 < r \leq R_2$.

It is instructive to present the local morphogenetic field 'w' in 3D version. Figures 3 - 6 show such a visualization of 'w' for selected values of parameter $p = 50$, 1, 0.5 and 0.3.

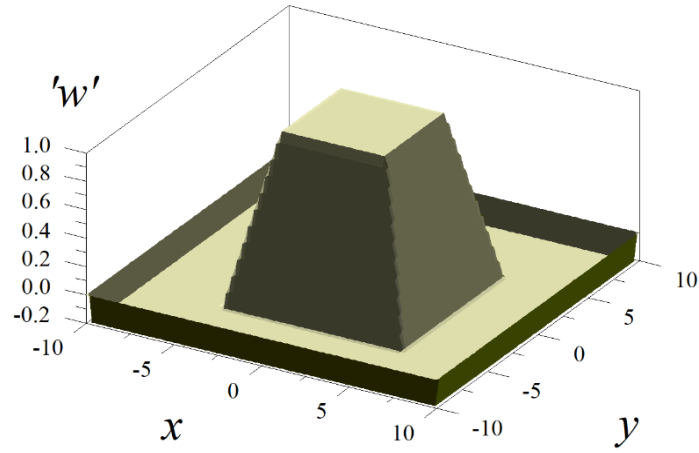

Fig. 3. A 3D visualization of the modified local activation-inhibition field for $p = 50$ and fixed values of the remaining parameters $w_1 = 1$, $w_2 = -0.2$, $R_1 = 4.8$, $R_2 = 10$ and $\Delta_1 = 1.8$. Note that $\Delta_2 \equiv \Delta_1 |w_2|/w_1$.

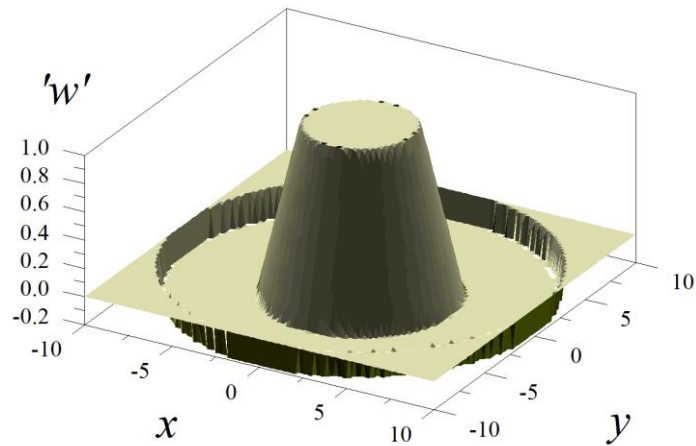

Fig. 4. The same as Fig. 3 but for $p = 1$.

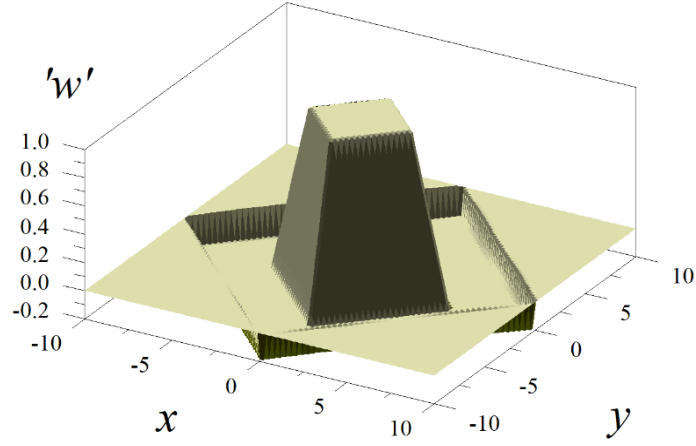

Fig. 5. The same as Fig. 3 but for $p = 0.5$.

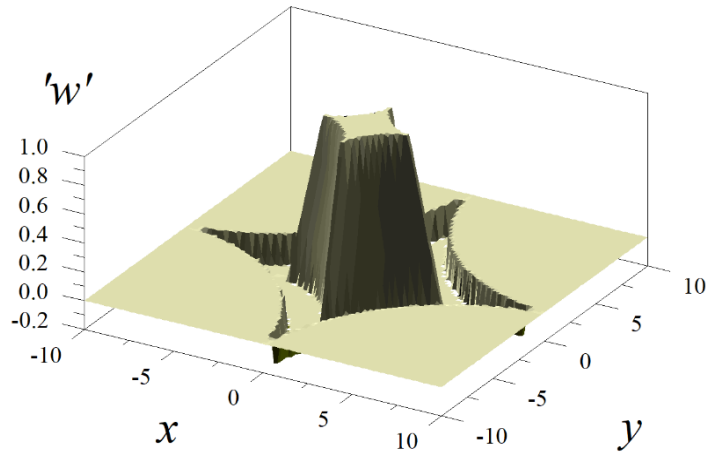

Fig. 6. The same as Fig. 3 but for $p = 0.3$.

Let us assume a neutral substrate material, *cf.* [9], which can be equally called an environment. In the current approach, the time evolution of a given cell $(x_0, y_0)$ depends on the local summary field $W(x_0, y_0; t)$, whose value is calculated as follows

$$W(x_0, y_0; t) = \sum_{r_i \in \mathbf{I}} w_1 \; + \; \sum_{r_i \in \mathbf{II}} w_{12}(r_i) \; + \; \sum_{r_i \in \mathbf{III}} w_2 \qquad (2)$$

Here, '*i*' refers to all DCs at positions $r_i$ in inner, transitional and outer regions. A morphogenetic field $W(x_0, y_0; t)$ is directly linked to the effective concentration of the two morphogens at that point and moment $t$.

Assuming in our model the existence of a certain threshold of the chemical reaction of cells, we extend the scope of its applicability. Therefore, we introduce the so-called inactive zone $[-W^*, W^*]$, where $W^*$ is the threshold value of the summary field. Only when the current $W(\mathrm{x}, \mathrm{y}) > W^*$ or $W(\mathrm{x}, \mathrm{y}) < -W^*$, the state of cell $(\mathrm{x}, \mathrm{y})$ may change according to the rules listed below. The temporal evolution of the pattern can be expected to be more stable compared to the previously considered no-threshold case [8, 9]. Fig. 7 shows a diagram illustrating the role of the threshold value of the local summary field $W$ for non-zero $W^*$ and for the opposite case $W^* = 0$.

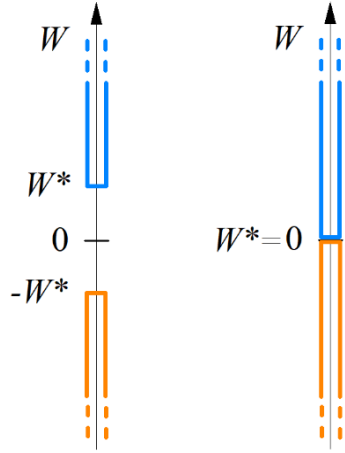

Fig. 7. The current modified Young's CA takes into account a passive zone with a width of $2W^*$ for the local summary field $W$ (left), while in original Young's model there is no passive zone (right).

Now, we recall the main rules for the modified model. The independent basic model parameters, $w_1$, $w_2$, $R_1$, $R_2$, $\Delta_1$ and $p$, relate to a local morphogenetic field presented in Fig. 1. When a non-zero threshold $W^*$ is also involved, the modified model allows considering a richer spectrum of cell behavior. One remark arises here. With a fixed threshold value, we expect that the smaller $p$, the relatively stronger the influence of the threshold on the evolution of the pattern. Presumably, this effect is caused by the decrease in the area of individual regions I, II and III when $p$ decreases.

For each grid cell $(x_0, y_0; t)$ in the next time step $t + 1$, the following situations are possible:

if $\quad W > W^* \quad$ then $\;$ UC (DC) becomes (remains) a DC, $\hfill$ (3a)

if $\quad -W^* \leq W \leq W^* \quad$ then $\;$ the considered cell does not change its state, $\hfill$ (3b)

if $\quad W < -W^* \quad$ then $\;$ DC (UC) becomes (remains) a UC. $\hfill$ (3c)

It is obvious that for $W^* = 0$ and in the absence of a transitional region, the current model simplifies to Young's CA for a neutral substrate material. When the results of the first state changes for each grid cell are recorded as a separate sequential pattern, this moment can be described as the first iteration step $j = 1$ (it can also be called the time evolution step $t = 1$). The total time of evolution can conveniently be measured in iterative steps. Then, the resulting binary pattern with a current DC-population of size $n(j)$ becomes the new initial configuration. Within this approach, cell updating can be treated as synchronous. Let us denote the number of "positive" UC → DC and "negative" DC → UC changes in the $j$-th iteration by $\Delta n^+(j)$ and $\Delta n^-(j)$. When $\Delta n^+(j) = \Delta n^-(j) = 0$, we stop repeating the iteration process. This means that the evolving system has reached a stationary, stable final configuration, and the pattern formation process itself can be considered stable.

## 3. Illustrative examples

Our approach can also be applied to cases where the initial configuration contains compact clusters with $p$-defined shapes instead of single DCs. Thus, we would like to begin with the synthetic configurations with given concentrations of black phase $\varphi_{init}$ on a square grid of linear size L = 81 pixels. The first initial configuration is *deterministic* quasi-Sierpinski (DSC) carpet for squared discs; *cf.* Fig. 8. The remaining initial configurations are *random* quasi-Sierpinski (RSC) carpets for squared (Fig. 9), circular (Fig. 10), rhombic (Fig. 11) and hypocycloid (Fig. 12) discs. For comparison, the final patterns for the corresponding random DC initial configurations are also shown in Figs. 9 - 12.

The chosen values of the independent model parameters are $w_1 = 1$, $w_2 = -0.3$, $R_1 = 4.8$, $R_2 = 10$, $\Delta_1 = 0.45$ and threshold $W^* = 0.48$. The auxiliary parameter $\Delta_2$ is not used in an explicit form because it can be expressed as $\Delta_2 \equiv \Delta_1 |w_2|/w_1$; see Fig. 1. Hence, with given $w_1$, $w_2$ and $\Delta_1$, the slope of the local field in the transition zone described as $w_1/\Delta_1$ is also established. Additionally, for selected patterns, we will show for the first time the $W$ surface illustrating the regional summary field in the final stage of evolution.

We will first focus on the role of different shapes of activation/inhibition regions as applied to initial synthetic configuration of compact clusters with pseudo fractal geometry. According to Eq. (1), for a large $p >> 1$, the shape of the activation/inhibition area is close to a square. Typically, in computer simulations we treat pixels as unit squares. Here, it is enough to use $p = 50$ to obtain activation/inhibition areas consistent with the square objects present in the initial configuration. In Fig. 8, starting with a given synthetic configuration, we compare the consistent final pattern ($p = 50$) with the selected inconsistent one ($p = 1$). In the latter case, the final DC concentration slightly decreases, while the number of iterations increases. However, in both cases the overall symmetry of the pattern is preserved. Taking the consistent case as an example, we show in Fig. 8(d) the relating final surface $W$. As can be seen, positive (negative) values correspond to the black (white) pixels in Fig. 8(b). Again, the general symmetry is preserved for the corresponding figures.

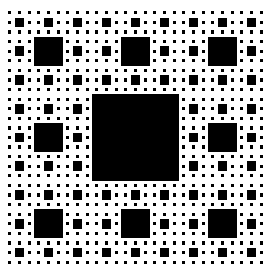 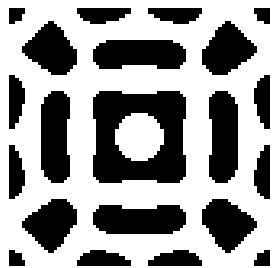 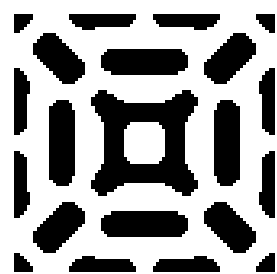

(a) $\varphi_{init} = 0.3757$   (b) $\varphi_{final} = 0.4323$   (c) $\varphi_{final}* = 0.4292$

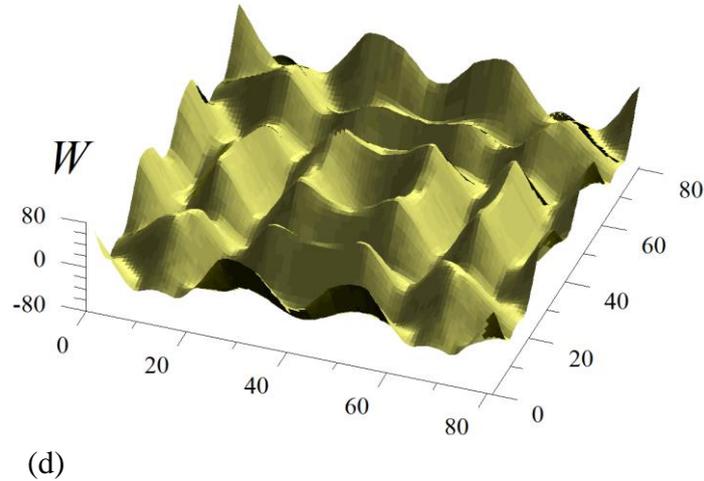

(d)

Fig. 8. Effect of geometry of activation/inhibition region on the patterning process. (a) The fourth iteration of a square-based inverted deterministic Sierpinski carpet (DSC) as selected initial synthetic configuration. (b) Corresponding final pattern obtained after $j = 25$ iterations using compatible geometry. (c) Final pattern obtained after $j = 42$ iterations, this time using inconsistent geometry for $p = 1$. (d) The explicit form of the final surface $W$ is presented for the consistent case. Positive (negative) values correspond to the black (white) pixels in (b).

In Figures 9 - 12, we consider initial configurations of a random Sierpinski carpet (RSC) type for the following values of the deformation parameter $p = 50$, 1, 0.5 and 0.3. For comparison purposes, for each of them we also create the initial configuration of randomly distributed DCs of the corresponding concentration. The latter initial configurations are not shown here.

Again, for each of the selected values of the deformation parameter $p$, the considered pairs of the final patterns show similar geometric features. Note that the final concentrations in pairs are higher for randomly distributed DCs compared to the RSC cases.

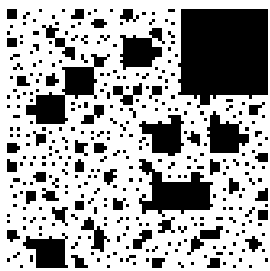

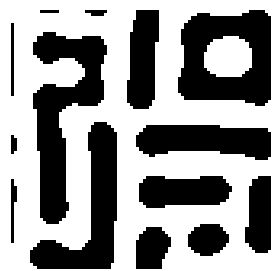

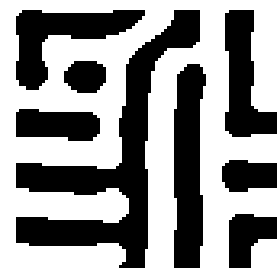

(a) $\varphi_{\text{init}} = 0.3757$    (b) $\varphi_{\text{final}} = 0.4588$    (c) $\varphi_{\text{final, rnd pix}} = 0.4781$

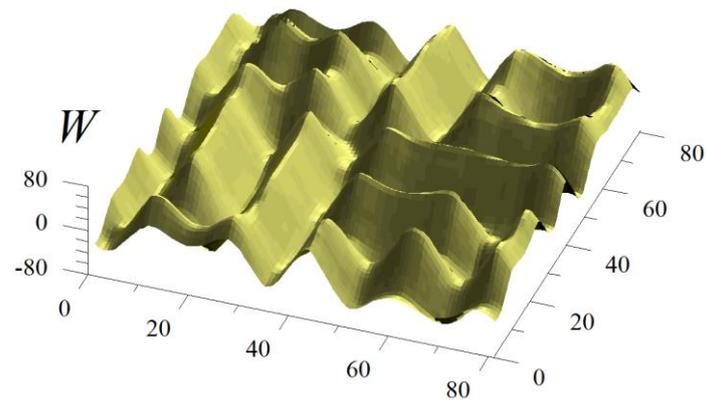

(d)

Fig. 9. Effect of different types of random initial configurations on the patterning process using appropriate geometry of activation/inhibition regions with $p = 50$. (a) The fourth iteration of a square-based inverted random Sierpinski carpet (RSC). (b) The corresponding stable final pattern after $j = 39$ iterations. (c) Related pattern for randomly distributed DCs with the same initial concentration after $j_{\text{rnd pix}} = 76$ iterations. (d) The final surface $W$ shown for illustration purposes refers to case (b).

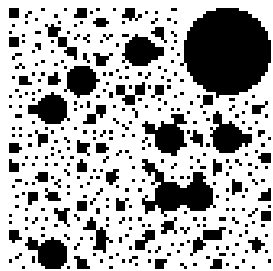 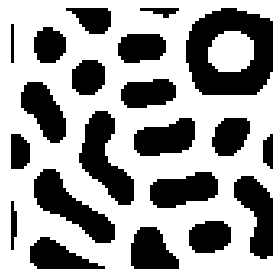 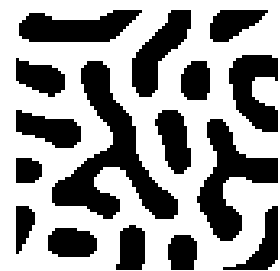

(a) $\varphi_{\text{init}} = 0.3379$     (b) $\varphi_{\text{final}} = 0.4193$     (c) $\varphi_{\text{final, rnd pix}} = 0.4441$

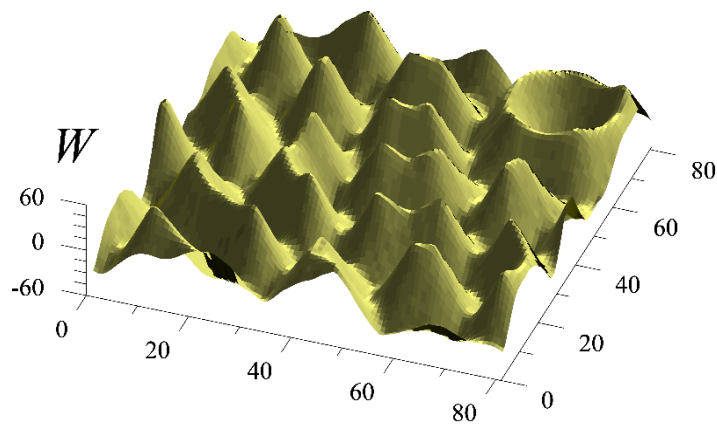

(d)

Fig. 10. As in Fig. 9 but with $p = 1$ for randomly distributed circular discs after $j = 26$ iterations and for randomly placed DCs of the same concentration after $j_{\text{rnd pix}} = 24$.

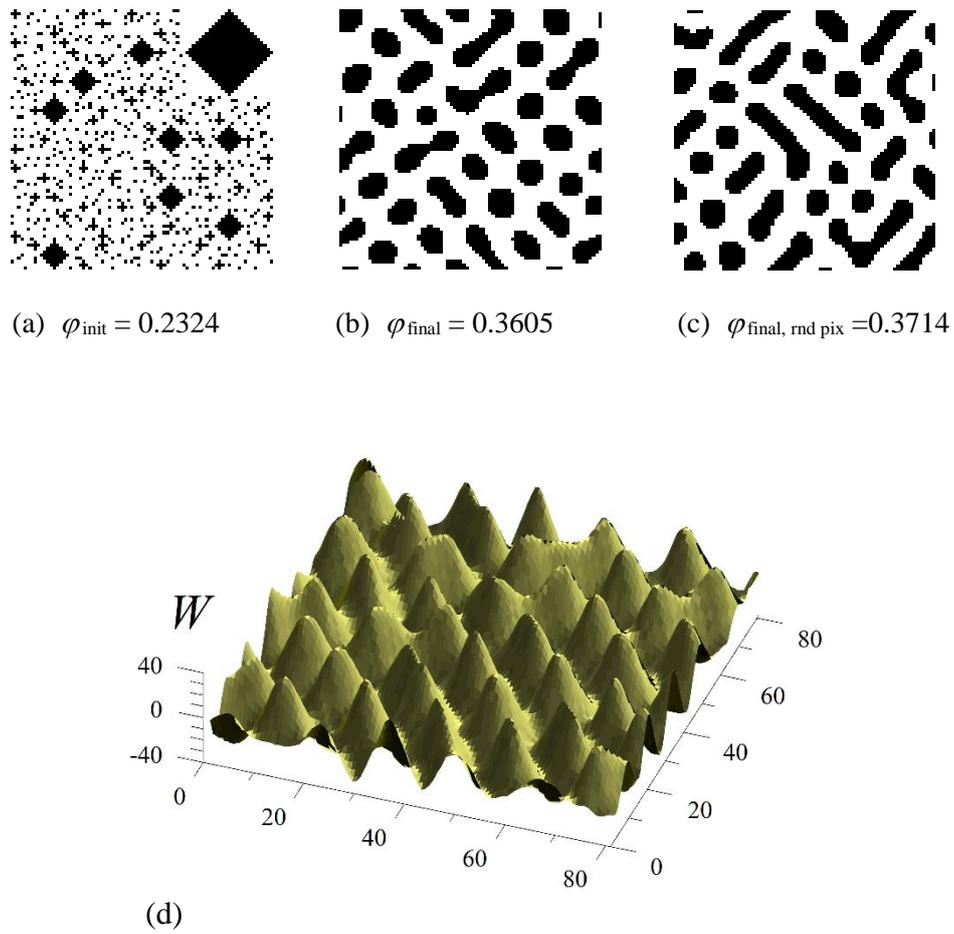

(a) $\varphi_{\text{init}} = 0.2324$     (b) $\varphi_{\text{final}} = 0.3605$     (c) $\varphi_{\text{final, rnd pix}} = 0.3714$

(d)

Fig. 11. As in Fig. 9 but with $p = 0.5$ for randomly distributed rhombic discs after $j = 15$ iterations and for randomly placed DCs of the same concentration after $j_{\text{rnd pix}} = 24$.

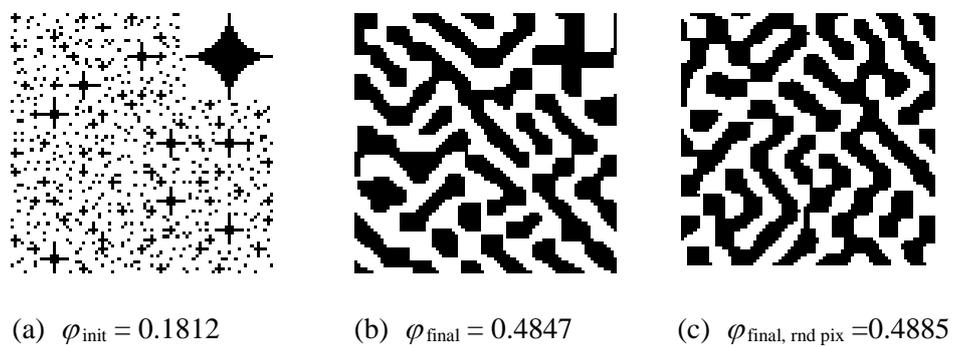

(a) $\varphi_{\text{init}} = 0.1812$     (b) $\varphi_{\text{final}} = 0.4847$     (c) $\varphi_{\text{final, rnd pix}} = 0.4885$

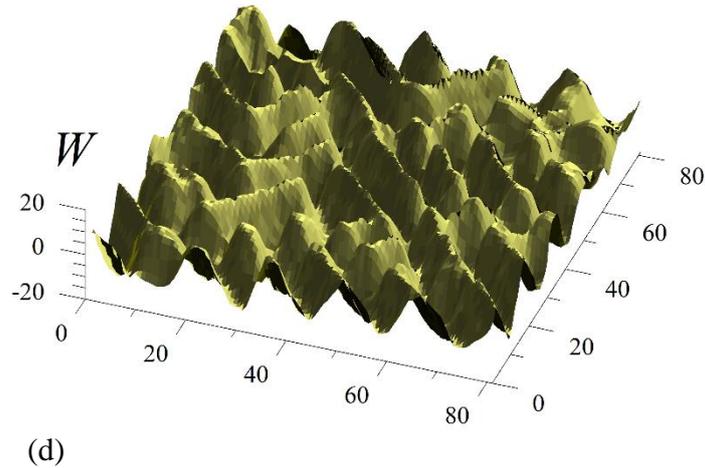

(d)

Fig. 12. As in Fig. 9 but with $p = 0.3$ for randomly distributed hypocycloid discs after $j = 23$ iterations and for randomly placed DCs of the same concentration after $j_{\text{rnd pix}} = 16$.

Additional information can be obtained by observing the spatiotemporal evolution of patterns formed for an identical random population of 5% DC. For this purpose, we use a larger size L = 128 pixels for subsequent patterns. However, we keep the same set of basic model parameters, that is $w_1 = 1$, $w_2 = -0.3$, $R_1 = 4.8$, $R_2 = 10$, $\Delta_1 = 0.45$ and threshold $W^* = 0.48$. Fig. 13 shows the evolution of the DC concentration during the pattern formation process for the previously selected $p$-collection.

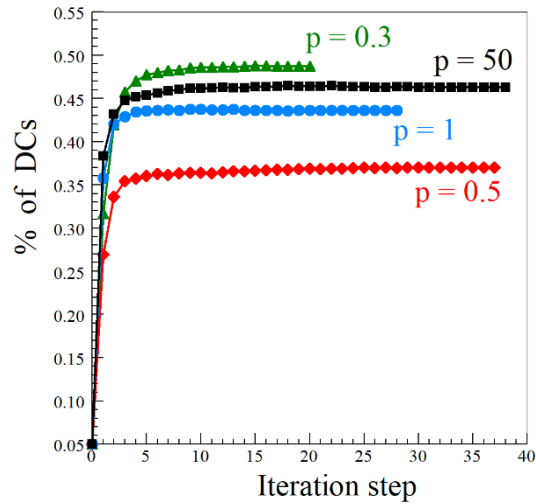

Fig. 13. Changes in DC concentration for $p$-patterns created from a given 5% random DC population on a square grid of linear size L = 128 pixels for selected values of $p = 50$, 1, 0.5, and 0.3. The corresponding symbols are squares, circles, rhombuses, and triangles, respectively.

It should be noted that the order of the final concentrations from the highest to the lowest, i.e. at $p = 0.3$, 50, 1 and 0.5, is the same as for the corresponding patterns in Figures 9 (c) – 12 (c). Such ordering occurs even though the above-mentioned patterns were created using different initial random configurations for the chosen concentrations. Interestingly, in both cases, the evolution time measured by the number of iteration steps is the shortest for the pattern with the highest final DC concentration with $p = 0.3$.

Even though the evolution of each pattern started from the same random population of DCs, which is by definition a statistically isotropic configuration, the final $p$-patterns, *cf.* insets in subsequent Figures 4 (a), (c) and (d), exhibit some features of biaxial anisotropy except for case (b). Since in this case $p = 1$, i.e. we have a circular activation/inhibition region that does not distinguish any direction, the lack of anisotropy in the final pattern is fully understandable. In principle, traces of biaxial anisotropy in the final patterns can be revealed using the orthogonal two-point correlation function $S_2$.

First, for each final pattern and $0 < r \leq L/2$, we observe small differences in the $S_2(r)$ function values computed separately for the two orthogonal directions consistent with the principal axes. Therefore, we use $S_2(r)$ averaged over these two directions and denote as $S_2(r; +) \equiv [S_2(r; |) + S_2(r; -)] / 2$. A similar effect appears on two perpendicular diagonals and for them we use a similar mean function, i.e. $S_2(r; \times) \equiv [S_2(r; \backslash) + S_2(r; /)] / 2$. In Figures 14 (a) – (d), we compare the functions $S_2(r; +)$ and $S_2(r; \times)$, which represent filled and open symbols, respectively. As mentioned above, in case (b) the final pattern should be statistically isotropic. Indeed, in the most interesting region of small distances, e.g. for $r \leq 30$, the $S_2(r; +)$ and $S_2(r; \times)$ curves are very similar to each other. As expected, the curves practically coincide in the region, while in each of the other cases, i.e. for (a), (c) and (d), the pairs of curves are directionally sensitive and differ significantly. Moreover, one can observe a characteristic shift of the first maximum of the $S_2(r; +)$ curve to the left, up to the exchange of positions with the first maximum of the $S_2(r; \times)$. Consistent with this observation is the decreasing average width of the black areas for these two orthogonal directions compared to the diagonal directions, see the corresponding insets.

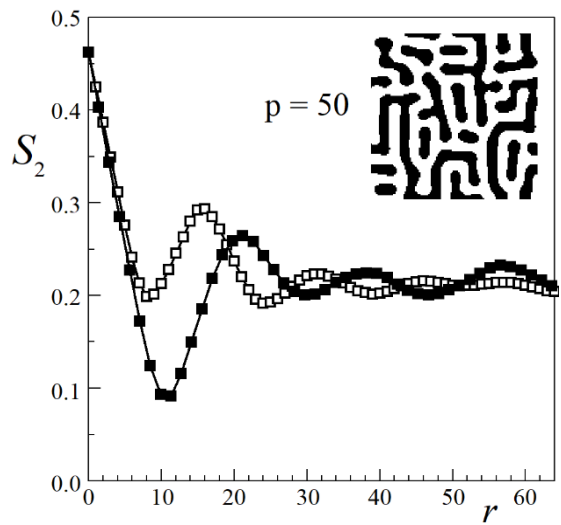

(a)

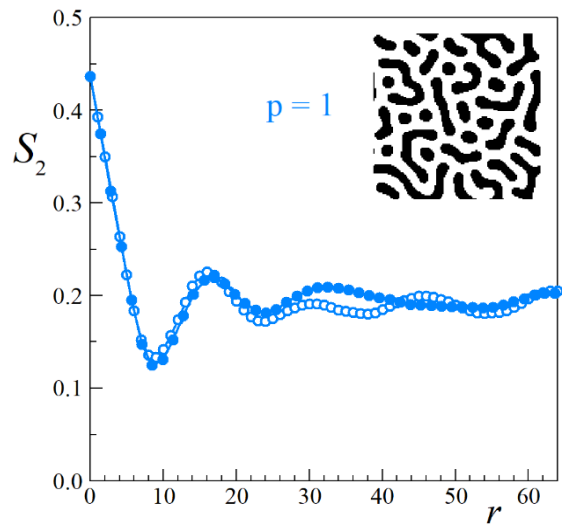

(b)

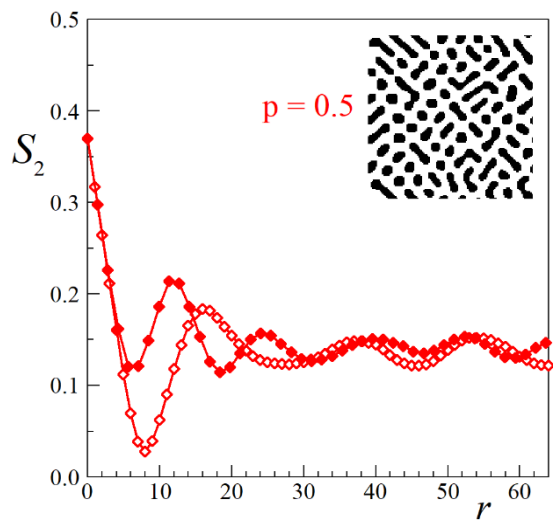

(c)

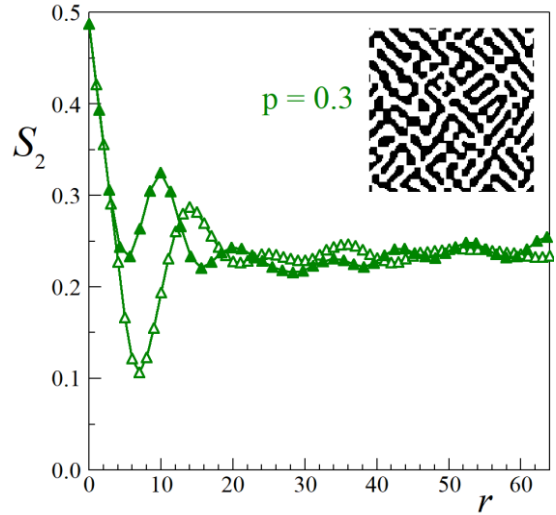

(d)

Fig. 14. The averaged orthogonal two-point correlation function computed for the stable final $p$-patterns shown in the insets. The filled symbols refer to the function averaged over two perpendicular directions consistent with the principal axes of the system and marked in the text as $S_2(r; +)$. Open symbols represent the function averaged over two perpendicular diagonals and denoted in the text as $S_2(r; \times)$. Notice that the first maximum of the $S_2(r; +)$ curve shifts to the left to swap positions with the first maximum of the $S_2(r; \times)$.

Finally, we consider the two-parameter dependence of the final DC concentration, $<\phi>$, averaged over ten different seeds. Recall that the first of the two control parameters, limited here to the range $-1 \leq w_2 < 0$, is a long-range inhibitor measuring the net strength of the inhibitory effect of the inhibitory effect in the outer region III. The second parameter, considered here in the range $0 < p \leq 1$ conveniently describes the shape of the activation/inhibition regions; see Eq. (1) and Fig. 2. The remaining model parameters are constant. These include two model parameters introduced for the first time in this work, $\Delta_1$ and $W^*$. The former, relating to the slope of the local field in the transition zone, is set here as $\Delta_1 = 1.8$. The other is the chemical signal strength threshold $W^*$. It is worth noticing that the higher the threshold $W^*$, the better the stability of the patterning process. Here, the value of $W^* = 0.48$ is sufficient for the percentage of all patterning processes that are stably completed to be greater than 80 percent. Furthermore, for most of them the number of iterations is less than 40. Using a step 0.01 for both control parameters, $w_2$ and $p$, we plot the overall surface in Fig. 15.

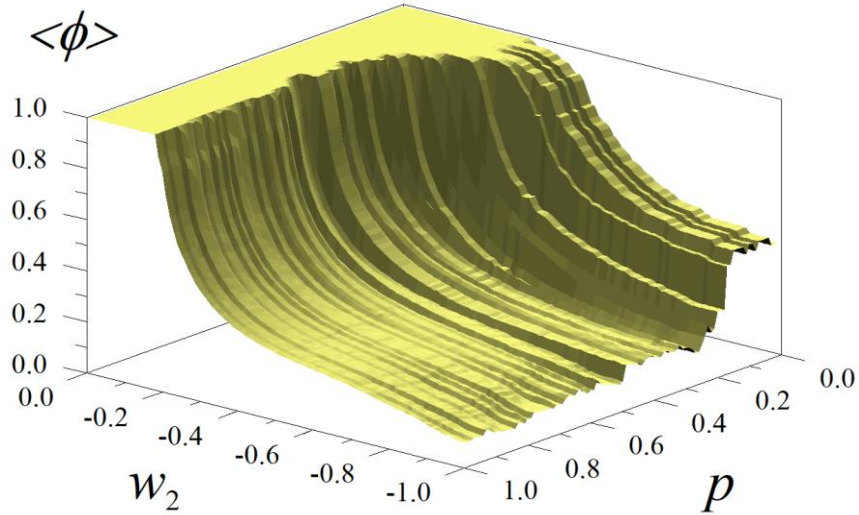

Fig. 15. Average final concentration as a function of the control parameters $w_2$ and $p$, both with a step 0.01. We recall that the other parameters are set to $w_1 = 1$, $R_1 = 4.8$, $R_2 = 10$, $\Delta_1 = 1.8$ and $W^* = 0.48$. Note that $\Delta_2 \equiv \Delta_1 |w_2|/w_1$.

In this way, for a given set of model parameters we can easily notice/recognize characteristic features of concentration evolution. It should be noted that the lines described by $<\phi> = f(w_2; p = \text{const.})$ and treated as a guide for the eyes, become smoother as $p$ increases in the interval considered here. When $p = 1$, the line corresponds to the isotropic case. On the other hand, especially for stronger anisotropy with decreasing $p$, the inhibitory effect of $w_2$ weakens, which is manifested by a higher final concentration. In this work, for illustrative purposes we focused on the influence of two basic control parameters on the mean final concentration. Of course, to perform a richer analysis, we need to consider different sets of model parameters. This should make it possible to find relationships between the values of the relevant model parameters and different types of anisotropic final patterns. It would also be interesting to find the possibility of applying a modified CA model with appropriately shaped non-circular activation/inhibition regions to directionally hyperuniform microstructures [14].

## 4. Concluding remarks

A simple method based on Young's approach is presented, which allows modeling the directional dependence of the diffusion rate of a chemical signal in a biaxially anisotropic environment. Among the modifications of the CA model, we can distinguish two most important ones: the introduction of a threshold for the required absolute chemical signal strength and the use of activation/inhibition regions consistent with a given biaxial anisotropy. Partial results on pattern generation and their spatiotemporal evolution are presented. It has been confirmed that the presence of the $W^*$ threshold improves the stability of the patterning process. It is worth noting that for the first time we have visualized the $W$ field for the final example patterns. Using appropriate pairs of two-point orthogonal correlation functions, we observe characteristic shifts of the first maxima of the corresponding curves, confirming their directional sensitivity to changes in biaxial anisotropy. The overall effect of the two basic control parameters, $w_2$ and $p$, on the mean final DC concentration was also obtained. This enables easy detection of characteristic features of concentration evolution. The proposed modifications to the CA model strengthen its physical foundations and extend its applications, especially for other anisotropic substrates.

**Acknowledgement** We would like to thank Krzysztof Smaga for his participation in performing some of the computer calculations.

<div align="center">REFERENCES</div>

[1] J.D. Murray, *Mathematical Biology II: Spatial Models and Biomedical Applications*, 3rd edn., Springer, Berlin 2003.

[2] A.M. Turing, *Phil. Trans. Roy. Soc. Lond. B* **327**, 37 (1952).

[3] A. Gierer, H. Meinhardt, *Kybernetik* **12**, 30 (1972).

[4] H. Meinhardt, *Models of Biological Pattern Formation*, Academic Press, London 1982.

[5] L. Edelstein-Keshet, *Mathematical Models in Biology*, SIAM, Philadelphia 2005.

[6] Y. Bar-Yam, *Dynamics of Complex Systems*, Perseus Books. New York, 1997.

[7] A. Ilachinski, *Cellular Automata. A Discrete Universe*, World Scientific, 2001.

[8] D.A. Young, *Math. Biosci.* **72**, 51 (1984).

[9] R. Piasecki, K. Smaga, W. Olchawa, D. Frączek, *Acta Phys. Pol. B* **49**, 961 (2018).

[10] L. Willis, A. Kabla, *J. R. Soc. Interface* **13**, 20151077 (2016).

[11] S. Kondo, *J. Theor. Biol.* **414**, 120 (2017).

[12] T. Ishida, *Micromachines* **9**, 339 (2018).

[13] Y. Jiao, F. H. Stillinger and S. Torquato, *Phys. Rev. Lett.* **100**, 245504 (2008).

[14] W. Shi, D. Keeney, D. Chen, Y. Jiao, and S. Torquato, *Phys. Rev. E* **108**, 045306 (2023).